\documentclass[12pt]{article}

\usepackage{scicite}
\usepackage{times}
\usepackage{soul}

\newenvironment{sciabstract}{%
\begin{quote} \bf}
{\end{quote}}

\usepackage{graphicx} % Graphiken einbinden: hier f"ur pdflatex 
\usepackage{epstopdf}
\usepackage{txfonts}
\usepackage{color}
\usepackage{float}
\usepackage{caption}

\DeclareCaptionLabelFormat{nospace}{#1#2} %removes spacing between S and number

\usepackage{soul}   %% use \st{}

\title{A whirling plane of satellite galaxies around Centaurus A challenges cold dark matter cosmology}

\author
{Oliver M\"uller$^{1\ast}$, Marcel S. Pawlowski$^{2}$, Helmut Jerjen$^{3}$, Federico Lelli$^{4}$\\
\\
\normalsize{$^{1}$Departement Physik, Universit\"at Basel, Klingelbergstrasse 82, CH-4056 Basel, Switzerland}\\
\normalsize{$^{2}$ {Department of Physics and Astronomy, University of California, Irvine, CA 92697, USA}}\\
\normalsize{$^{3}$ Research School of Astronomy and Astrophysics, Australian National University, Canberra,}\\
\normalsize{ACT 2611, Australia}\\
\normalsize{$^{4}$ {ESO - European Southern Observatory, Karl-Schwarschild-Strasse 1,}}\\ 
\normalsize{{85748 Garching bei M\"unchen, Germany}}
\\
\normalsize{$^\ast$To whom correspondence should be addressed; E-mail:  oliver89.mueller@unibas.ch.}\\
}

\date{}

\begin{document} 


\baselineskip24pt


\maketitle 

\begin{sciabstract}
The Milky Way and Andromeda galaxy are each surrounded by a thin plane of satellite galaxies that may be corotating. Cosmological simulations predict that most satellite galaxy systems are close to isotropic with random motions, so those two well-studied systems are often interpreted as rare statistical outliers. We test this assumption using the kinematics of satellite galaxies around the Centaurus\,A galaxy. Our statistical analysis reveals evidence for corotation in a narrow plane: of the 16  Centaurus\,A's satellites with kinematic data, 14 follow a coherent velocity pattern aligned with the long axis of their spatial distribution. In standard cosmology simulations, $<$ 0.5\% of Centaurus\,A-like systems show such behavior. Corotating satellite systems may be common in the Universe, challenging small-scale structure formation in the prevailing cosmological paradigm.

\end{sciabstract}

%________________________________________________________________

The presence of planes of satellite dwarf galaxies around the Milky Way \cite{1976RGOB..182..241K, 1976MNRAS.174..695L,2012MNRAS.423.1109P,2013MNRAS.435.1928P} and Andromeda \cite{2006MNRAS.365..902M,2006AJ....131.1405K,2013Natur.493...62I} galaxies have challenged our understanding of structure formation on galactic and subgalactic scales. Similar structures are rare in galaxy formation simulations based on the standard Lambda Cold Dark Matter ($\Lambda$CDM) cosmological model, which predicts close to isotropic distributions and random kinematics for satellite systems \cite{2005A&A...431..517K}. 
The existence of planes of satellite galaxies around these two largest galaxies in the Local Group is difficult to explain within the $\Lambda$CDM framework. Some authors have argued that preferential accretion of satellites along filaments may explain such flattened structures \cite{2005ApJ...629..219Z}. Others suggest that the Local Group should be considered a rare exception in an otherwise successful cosmological model \cite{1995Natur.377..600O,1993Natur.366..429W,2005ApJ...633..560E}.
This interpretation, however, has been challenged by emerging evidence for anisotropic satellite distributions around massive galaxies beyond the Local Group \cite{2017A&A...602A.119M, 2013AJ....146..126C}.

The cosmic expansion of the Local Void (a vast, empty region of space adjacent to the Local Group) has been suggested  as a possible origin for the formation of these planar structures \cite{2015MNRAS.452.1052L}. An issue which is mostly ignored in this context is the coherent 
kinematics of the satellite galaxies, which are likely corotating  around their host. This is clear for the Milky Way \cite{2013MNRAS.435.2116P, 2015MNRAS.453.1047P} where accurate proper motions are available for several satellites, but it remains more uncertain for Andromeda \cite{2013Natur.493...62I} because only velocities projected along the line of sight (LoS)  are measurable.
Such orderly kinematic motions are extremely rare in high-resolution cosmological N-body simulations \cite{2014MNRAS.442.2362P} and statistically should not be observed in typical galaxy groups. It remains unclear whether such planes of satellites are unique to the Local Group, or ubiquitous in the nearby Universe.

In this Research Article, we study the galaxy group in the constellation Centaurus. The Centaurus Group is the richest assembly of galaxies within a distance of 10\,megaparsecs (Mpc) from the Milky Way, the so-called Local Volume \cite{2004AJ....127.2031K,2013AJ....145..101K}. It comprises two concentrations: the Cen\,A subgroup dominated by a radio-active elliptical galaxy Centaurus\,A (Cen\,A, NGC\,5128) at a distance of 3.8\,Mpc, and the M\,83 subgroup dominated by a late-type spiral galaxy M\,83 (NGC\,5236) at a distance of 4.9\,Mpc \cite{2004AJ....127.2031K,2013AJ....145..101K}. The galaxies which are  gravitationally bound to Cen\,A  were claimed to be distributed in two parallel planes \cite{2015ApJ...802L..25T}. The discovery of additional satellite galaxies in the group weakened the case for a double-planar structure, whilst a single-plane interpretation has become more statistically significant \cite{2015A&A...583A..79M,2017A&A...597A...7M,2016ApJ...823...19C}. This plane has a small scale  height with a root-mean-square (rms) thickness of 69\,kiloparsecs (kpc) and a major axis rms length of 309\,kpc \cite{2016A&A...595A.119M}. We investigate the kinematics of this planar structure and compare it with galaxy formation simulations in $\Lambda$CDM cosmology. 

   \begin{figure*}
   \centering
   \includegraphics[width=16cm]{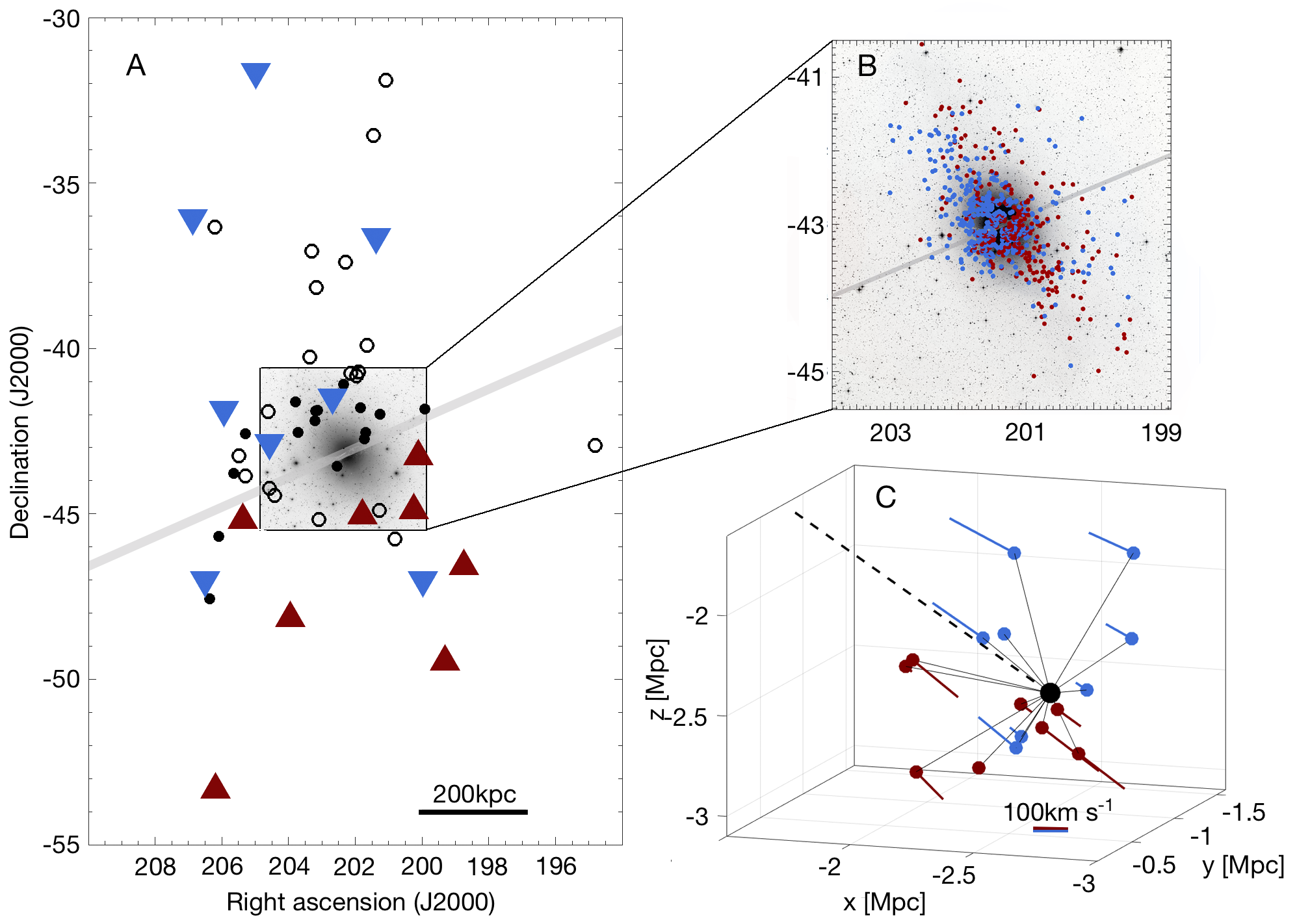}
      \caption{\textbf{On-sky and 3D distribution of the satellite system.} A: The on-sky distribution of the Cen\,A subgroup.  The central image of Cen\,A has been scaled up by a factor of five to illustrate the features of the host galaxy. Blue downwards and red upwards pointing triangles show approaching and receding satellite galaxies with respect to Cen\,A velocity, respectively.  Open circles are group member candidates, filled circles are confirmed satellites without velocity measurements. The line that optimally separates the approaching/receding satellites is indicated with the wide grey band; it coincides with the dust lane of Cen\,A.
{B: The kinematic distribution of 1239 planetary nebulae \cite{2015A&A...574A.109W}. Blue PNs are approaching, red PNs are receding relative to Cen\,A.} 
C: 3D representation of the spatial distribution of the Cen\,A satellite galaxies in equatorial Cartesian coordinates (the Earth is at $x=y=z=0$). The length of the colored lines is proportional to the observed velocity, the dashed line is our line of sight towards Cen\,A.}
\label{coor}
\end{figure*}
\section*{Dynamics of the Cen\,A satellite system}

From Earth the satellite plane around Cen\,A is seen nearly edge-on at an inclination of 14.6 degrees \cite{2016A&A...595A.119M}. This coincidental geometrical alignment allows us to scrutinize the kinematics of the plane. We use all available heliocentric velocities for the Cen\,A satellites, taken from the Local Volume catalog  \cite{2004AJ....127.2031K,2013AJ....145..101K}. The vast majority of satellites have accurate distances derived from the tip magnitude of the red giant branch (TRGB) method with a typical uncertainty of $\approx 5$\,percent. There are 31 confirmed satellites of Cen\,A with accurate distance measurements. Half of them have measured LoS velocities. One sample galaxy (KKs\,59) has a measured velocity but lacks a TRGB distance: we adopt the same distance as for Cen\,A; excluding this galaxy does not change our results. The adopted data are listed in Table S\ref{table:sample}.

The on-sky distribution of the satellites is plotted in Fig.\,\ref{coor} together with their motions relative to Cen\,A. Figure 1 also shows the positions and kinematic information for 1239 planetary nebulae \cite{2015A&A...574A.109W} and the 3D distribution of the satellites with measured velocities.
The mean velocity of the Cen\,A satellite system (555\,km\,s$^{-1}$) is equal to the recession velocity of Cen\,A (556$\pm$10\,km\,s$^{-1}$) within the measurement uncertainties. Hereafter, the recession velocity of Cen\,A is used as a zero-point reference and the terms approaching/receding are intended with respect to this velocity. The dust lane of Cen\,A serves as a natural dividing line: its position angle ($PA=110^{\circ}$) roughly coincides with the geometrical minor axis of the satellite plane \cite{2016A&A...595A.119M}. Clearly, approaching and receding satellites tend to lie to the South-West and North-East of the dividing line, respectively, indicating a kinematically coherent structure.

To determine the statistical significance of the kinematic coherence, 
we compare the velocities of Cen\,A satellites to a random phase-space distribution. Every galaxy has a 50\% chance of approaching or receding along the LoS. The probability of finding at least 14 out of 16 galaxies with coherent velocity movement is 0.42\%. Consequently, the observed velocity pattern of the Cen\,A satellites is statistically different from a random phase-space distribution at the 2.6$\sigma$ confidence level. 

   \begin{figure*}
   \includegraphics[width=8.5cm]{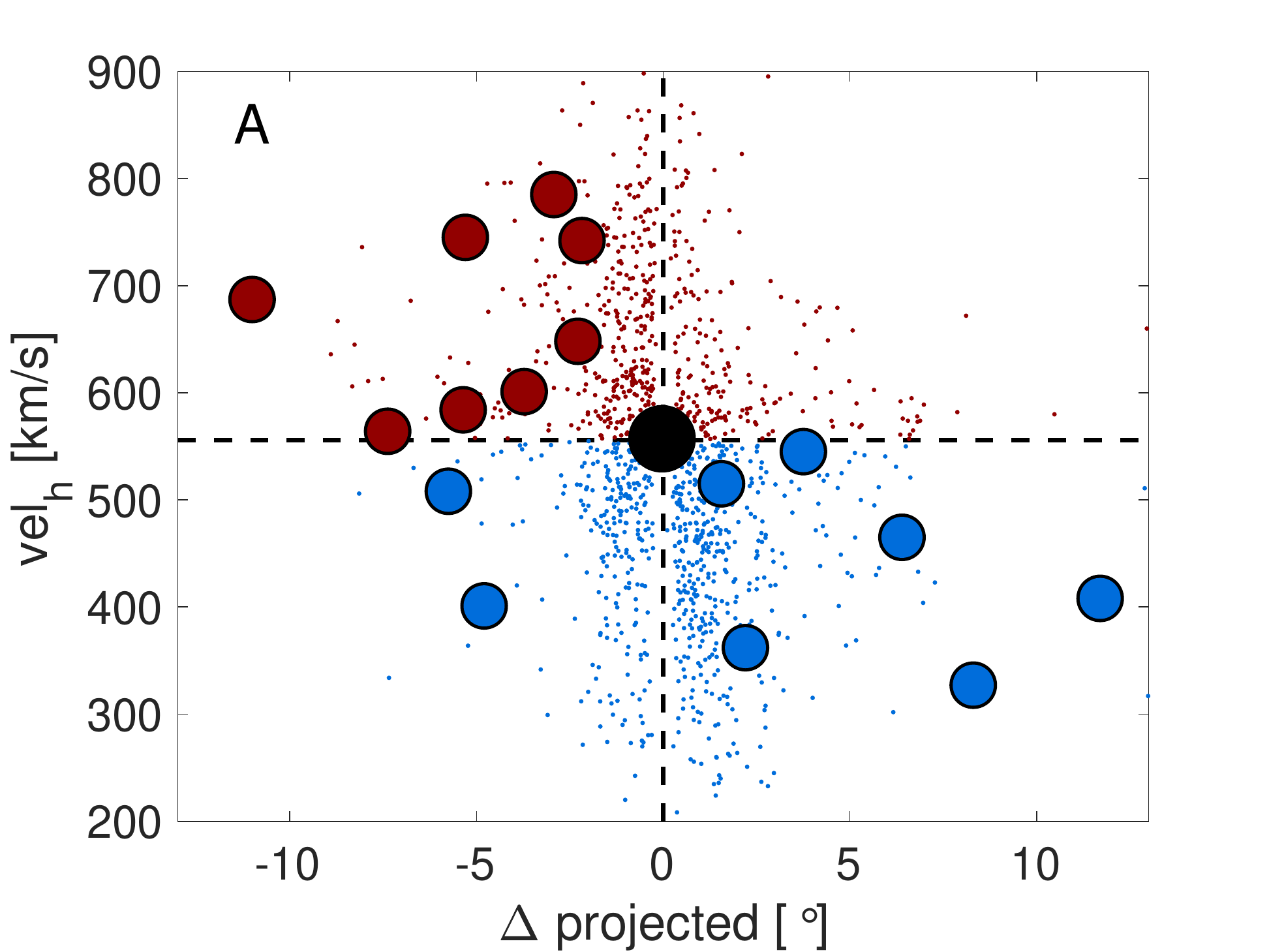}\hspace{-5mm}
   \includegraphics[width=8.5cm]{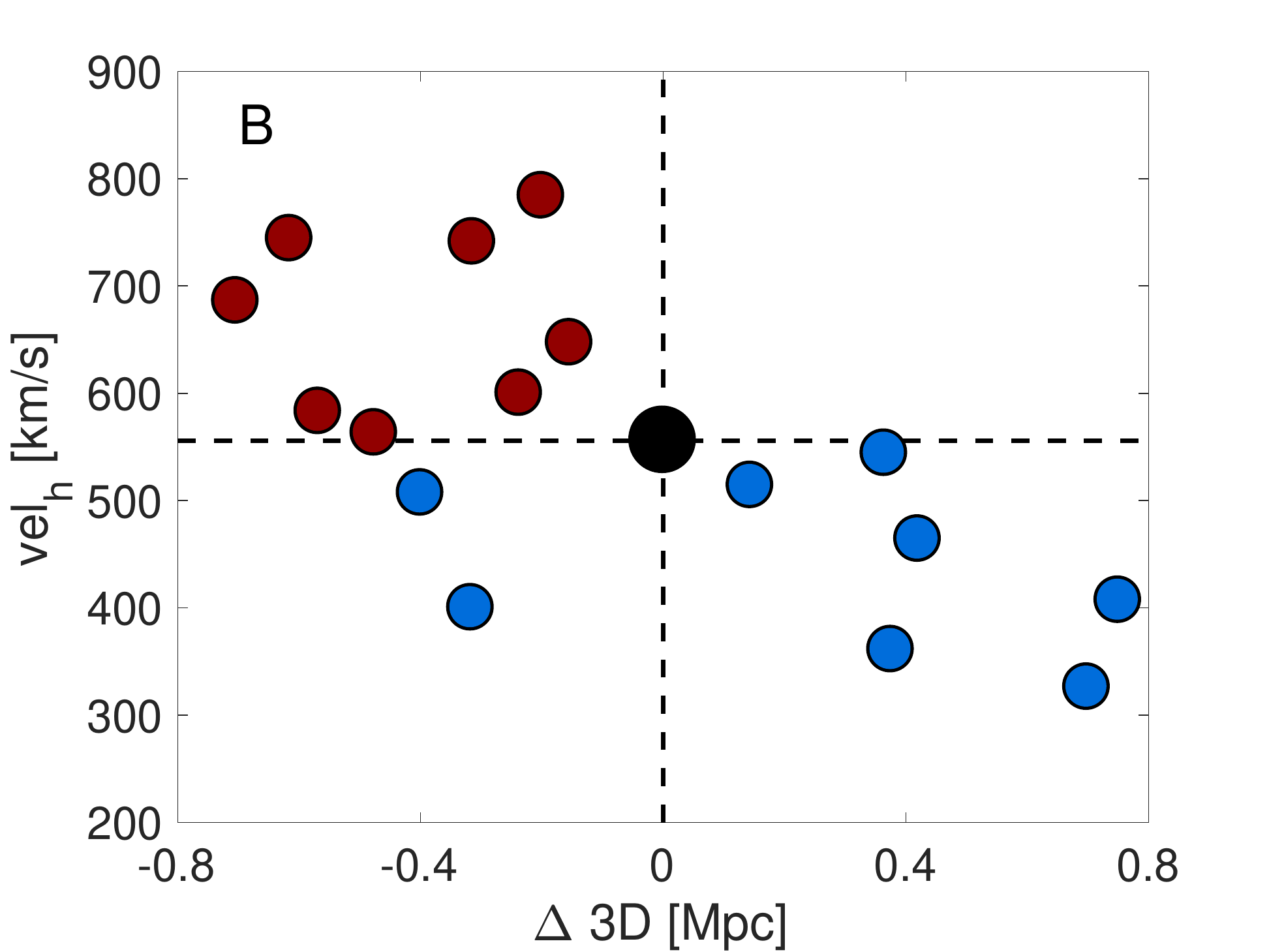}
      \caption{\textbf{Velocities and separations to Cen\,A.} Heliocentric velocities versus angular (A) and 3D (B) distances from Cen\,A (black dot), in the North (positive $\Delta$) or South (negative $\Delta$) of the dust lane. Large and small dots show, respectively, satellite galaxies and planetary nebulae. Blue and red colors indicate, respectively, approaching and receding objects with respect to the Cen\,A velocity. 
The angular distances of the PNs are scaled up by a factor of ten.}
         \label{velsep}
   \end{figure*}
Figure\,\ref{velsep} shows the heliocentric velocities of the satellite galaxies as a function of their distances to Cen\,A. The geometrical minor axis of the plane (or equivalently the dust lane) is used to assign a positive or negative sign to the distance between satellite galaxies and Cen\,A. Figure\,\ref{velsep} shows a clear trend: galaxies to the South of Cen\,A are approaching, whereas galaxies to the North are receding. This is to be expected if the satellites are rotating around Cen\,A. Only two satellite galaxies (KK\,221 and ESO\,269-058) deviate from this trend and may potentially be counter-rotating, analogous to the Sculptor dwarf in the Milky Way halo \cite{2011A&A...532A.118P}. An inspection of their properties and alignment inside the plane does not reveal any peculiar characteristics (e.g., they are not more massive or luminous than other satellites). The velocity field of the planetary nebulae within Cen\,A follows a similar trend: planetary nebulae in the northern and southern hemisphere are (on average) systematically blue and red shifted, respectively.

To explore the observed velocity pattern for the satellite galaxies, we ran three statistical tests, namely Pearson's R, Spearman's Rho, Kendall's Tau. These are standard methods to test correlations between independent variables. While the Pearsons's method tests for a strictly linear correlation, the Spearman's Rho and Kendall's Tau methods test for a general correlation between the variables. The null hypotheses is that velocities and separations are uncorrelated. The velocity pattern is significant within a confidence interval of 2$\sigma$ ($p$-value $<$ 0.03) for the projected separation and 3$\sigma$ ($p$-value $<$ 0.01) for the 3D separation \cite{Supp}.
These low p-values lead us to reject the null hypothesis,
implying a small chance of finding such a correlation in random, normal distributed data. We further consider how much more likely the hypothesis of correlated data is in respect to the hypothesis of uncorrelated data.  We applied a Bayesian correlation test \cite{Wetzels2012} and found that the scenario of coherently moving satellites is 4.5 times more likely using the projected separation and 16.5 times more likely using the full 3D information than uncorrelated satellite movements \cite{Supp}. Projected separations consistently give lower statistical significance than 3D distances because they contain less physical information: this highlights the importance of having TRGB distance measurements for dwarf galaxies in Centaurus.

\section*{Implications for galaxy formation}
The satellite galaxies in the Cen\,A subgroup collectively form a coherent kinematical structure. Comparable structures have been discovered in the Milky Way halo, where the majority of the 11 classical satellites share a coherent orbital motion (established with proper motion measurements of individual stars from the satellites) \cite{2013MNRAS.435.2116P}, and for the Andromeda galaxy, for which 13 out of 15 satellites follow a coherent LoS velocity trend \cite{2013Natur.493...62I}.

While we find that the kinematics of the Cen\,A satellites are unlikely to occur by chance, this does not immediately allow us to draw conclusions about its agreement with predictions from $\Lambda$CDM cosmology. Satellite galaxy systems in cosmological simulations generally exhibit some degree of phase-space coherence, due to the accretion of sub-halos from preferred directions, along filaments and in groups \cite{2005ApJ...629..219Z}. To judge whether this effect is sufficient to explain the observed coherence in the Cen\,A satellites, we determined the occurrence of such extreme structures in two cosmological simulations: Millennium II \cite{2009MNRAS.398.1150B} and Illustris \cite{2014Natur.509..177V}. Millennium\,II is a dark-matter-only $N$-body simulation that includes gravitational effects such as sub-halo accretion from filaments, but neglects baryonic effects  such as stellar and black-hole feedback and possible destruction of satellite galaxies due to the enhanced tidal effects from the baryonic disk \cite{2017MNRAS.471.1709G}. The relative importance of these effects is highly debated \cite{2016MNRAS.460.4348B,2015ApJ...800...34G,2014MNRAS.438.2916B,2017ARA&A..55..343B}. Hence, we also analyze the hydrodynamical Illustris simulation \cite{2014Natur.509..177V}, which additionally includes gas physics, star formation, and feedback processes.

\begin{figure}[H]
   %\centering
   \includegraphics[width=7.9cm]{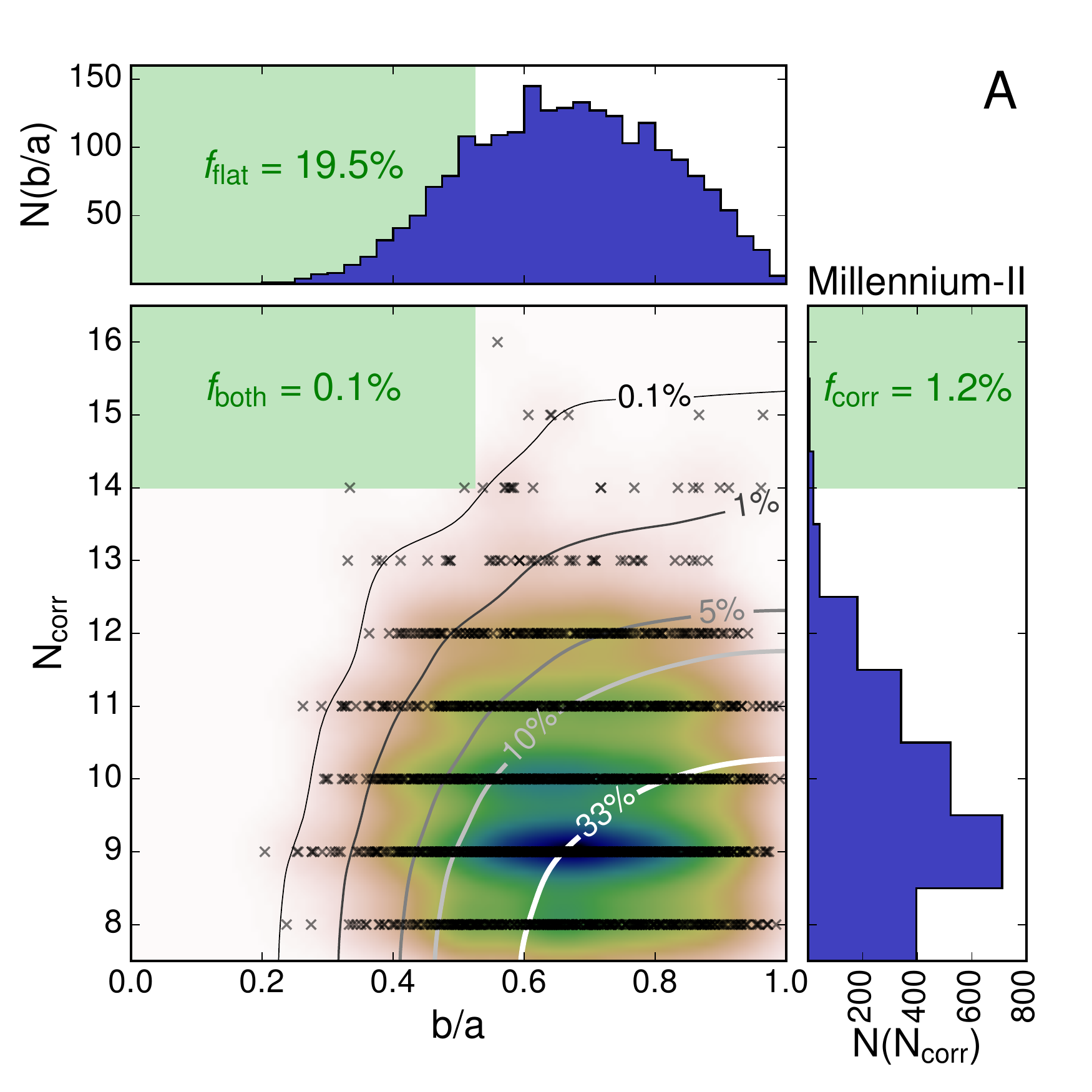}
   \includegraphics[width=7.9cm]{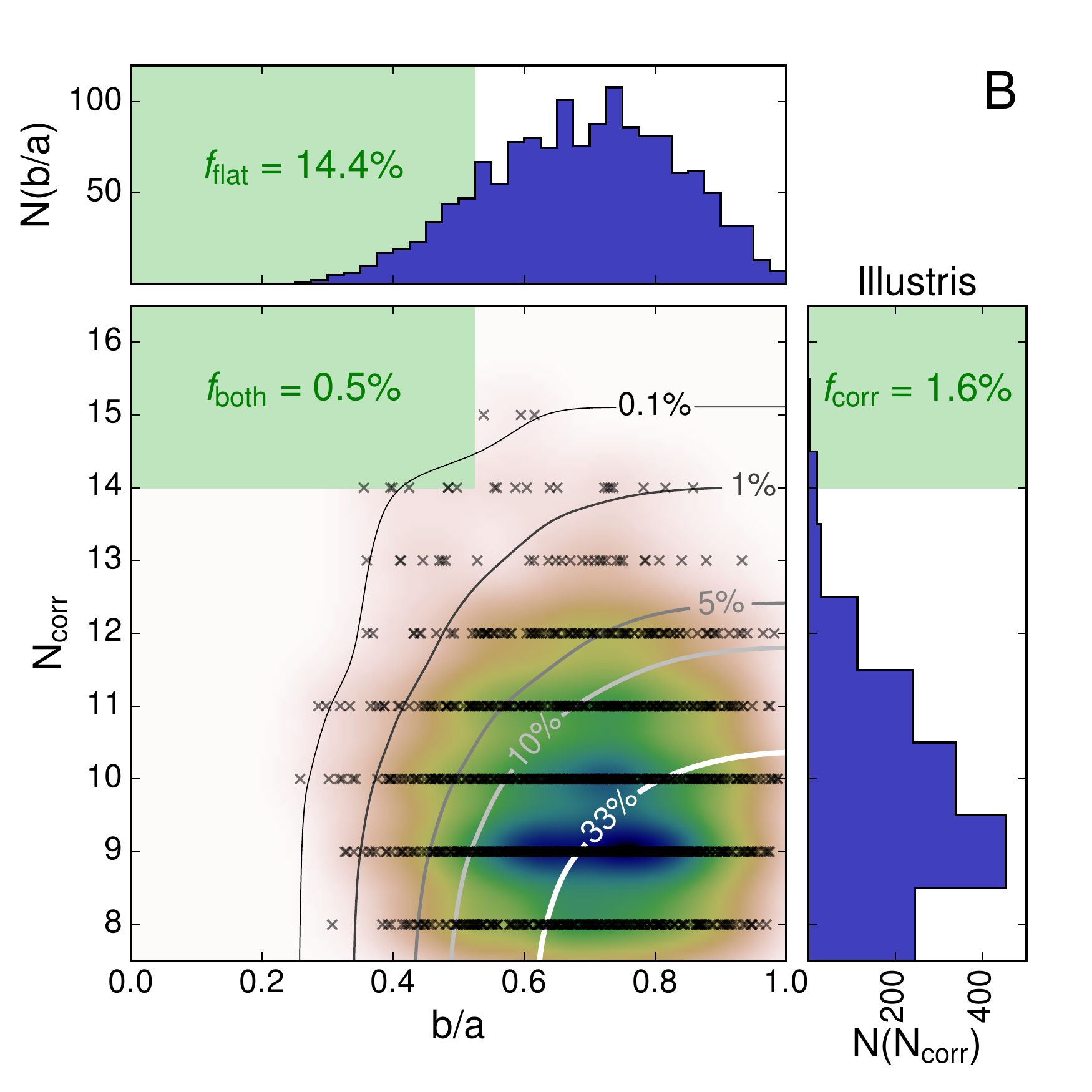}
   \caption{
   \textbf{Comparison to $\Lambda$CDM cosmological simulations.} The number of kinematically correlated  satellites $N_\mathrm{corr}$\ and the on-sky axis-ratio flattening $b/a$\ are plotted for Cen\,A analogs from the Millennium\,II (panel A) and Illustris (panel B) simulations. The density of simulated systems is indicated by the color map. The contours show the frequency of more extreme realizations, i.e., the number of points to the upper left of each position divided by the total number of realizations. The top and right histograms show the number of realizations with a given axis ratio $N(b/a)$ and a given number of correlated velocities $N(N_\mathrm{corr})$, respectively. The green areas delineate the regions in which systems are as or more extreme than the observed one: $f_{\mathrm{flat}}$, $f_{\mathrm{corr}}$, and $f_{\mathrm{both}}$\ give the frequency of realizations that are, respectively, at least as flattened as the observed system, at least as kinematically correlated, or both. A system has to fulfill both conditions simultaneously to reproduce the observed properties of the Cen\,A satellite system.}
         \label{fig:sims}
   \end{figure}

Our approach is analogous to recent studies of the frequency of the satellite planes around the Milky Way and the Andromeda galaxy \cite{2014MNRAS.442.2362P,2015MNRAS.452.3838C}. We identify Cen\,A analogs  within the simulations by selecting dark matter halos with masses between $4-12 \times 10^{12}$ solar masses ($M_{\odot}$) and by rejecting any candidate hosts that have  a companion galaxy with dark matter halo mass $\geq 1 \times 10^{12} M_{\odot}$\ within 1.4\,Mpc distance. We require a simulated galaxy-satellite system to fulfill two simplified criteria to be considered similar to the observed system: (i) the projected on-sky axis ratio of the system must be $b/a \leq 0.52$, where $a$ and $b$ are the semi-major and semi-minor axes, respectively, and (ii) the kinematic coherence along the long axis is at least 14 out of 16 satellites. We find that the occurrence of arrangements similar to Cen\,A in the cosmological simulations is 0.1 per cent for Millennium\,II and 0.5 per cent for Illustris (Fig.\,\ref{fig:sims}). These estimates must be considered upper limits, since we do not take into account the full 3D distribution of satellite galaxies. Even though the hydrodynamical Illustris simulation does contain a higher frequency of systems analogous to Cen\,A than the dark-matter-only Millennium\,II simulation, they are  rare cases in both.
The observed Cen\,A satellite system is thus in serious tension with the expectations from these $\Lambda$CDM simulations, to a similar degree as the satellite planes in the Local Group.

Could the coherent motion be the result of cosmic expansion? If that were the case, a correlation between the velocities of the satellites and their distances to the Milky Way would be expected. This is not found for the sample of Cen\,A satellite galaxies (Figure\,\ref{coor}, C). We thus can rule out that the cosmic expansion is responsible for the observed velocity field. Another possible origin of a velocity gradient is a perspective effect. For angular offsets $\delta$\ along the direction of motion, a fraction $\sin(\delta)$\ of a system's bulk tangential velocity is projected along the line of sight \cite{2008ApJ...678..187V,2016MNRAS.456.4432S}. The velocity gradient found for the Cen\,A system in Fig.\ref{velsep} implies a tangential velocity of the order of 1,000\,km\,s$^{-1}$, comparable to what would be required for the Andromeda satellite plane \cite{2013Natur.493...62I}.  This is unphysically high given that it exceeds the cosmic expansion at the distance of Cen\,A by a factor of 3.7. Such an interpretation of the velocity gradient would imply that the Cen\,A group moves in the direction defined by the satellite plane, which is unlikely. Another potential systematic issue is the contribution to the LoS velocity by the motion of the Sun around the Galactic centre. However, we found that this contribution is negligible: $-2$ to $-4$\,km\,s$^{-1}$, depending on the sky position, well within the uncertainties of the heliocentic velocity measurements (Table S1).

The coherent kinematics of the Cen\,A satellites, instead, is best explained by corotation within the plane. We explored a toy model with purely circular orbits and tried to deproject the LoS velocities into circular velocities \cite{Supp}. The results were unsatisfactory as many satellites would have unrealistic circular velocities, which randomly vary with the distance from Cen\,A.  This  suggests that the satellites must be on elliptical orbits, as expected for collisionless objects. Two galaxies do not follow the general trend: they may be counter-rotating, or on highly elliptical orbits, or simply unrelated to the planar structure. PNe provide additional evidence \cite{2004ApJ...602..685P}: they also show  coherent motion albeit this is less pronounced than for the dwarf satellites, only 65\% of PNe partake in the common motion (Figure 1).  Since the same trend is present in two independent populations of objects with different orbital times, we can expect this correlation to be long lived and thus indicative of corotation within the planes.

Corotation outside the Local Group has been investigated using satellite galaxy pairs on opposite sides of their hosts \cite{2014Natur.511..563I}.
The LoS velocities of satellite pairs are preferentially anti-correlated, suggesting a high incidence ($>50\%$) of corotating satellite pairs in the Universe \cite{2014Natur.511..563I}, although that result remains controversial \cite{2015MNRAS.453.3839P,2015MNRAS.449.2576C,2015ApJ...805...67I}. 
For the Cen A subgroup, the presence of a plane of satellite galaxies is known independently of velocity information and is established using multiple group members.  This is unlike previous studies, which were unable to determine whether specific pairs of satellites actually lie in a plane \cite{2014Natur.511..563I}.

In alternative frameworks for the formation of dwarf galaxies, co-rotating planes of satellites could be a consequence of past interactions and mergers between disk galaxies \cite{2013MNRAS.435.2116P}. During galaxy mergers, tidal tails form from disk material due to angular momentum conservation and can collapse into tidal dwarf galaxies \cite{2001AJ....122.2969H, 2007Sci...316.1166B, 2015A&A...584A.113L}. Hydrodynamical simulations show these may survive the interaction and begin orbiting around the central merger remnant as dwarf satellites \cite{2006A&A...456..481B, 2015MNRAS.447.2512P}. In the Local Group, a major merger forming the Andromeda galaxy has been proposed as a possible origin of the observed satellite galaxy planes around both the Milky Way \cite{2012MNRAS.427.1769F} and the Andromeda galaxy \cite{2013MNRAS.431.3543H}. The recent finding of a correlation between the size of spiral galaxy bulges (thought to form via major mergers) and the number of satellites is in agreement with this picture \cite{2016ApJ...817...75L}. Even the existence of some counter-rotating satellites can be understood in this framework \cite{2011A&A...532A.118P}. 

In summary, we find evidence for a kinematically coherent plane of satellite galaxies around Cen\,A, demonstrating that the phenomenon is not restricted to the Milky Way and Andromeda galaxies The kinematic coherence can be understood if the satellites are co-rotating within the plane, as seen around the Milky Way.
Considering that the likelihood of finding a single kinematically coherent plane is $\lesssim$0.5\% in cosmological $\Lambda$CDM simulations, finding three such systems in the nearby Universe seems extremely unlikely.

\section*{Acknowledgments}
The authors thank Bruno Binggeli and Marina Rejkuba for interesting discussions and helpful inputs. O.M. thanks Eva Schnider for inputs concerning the figures. O.M. is grateful to the Swiss National Science Foundation for financial support. 
M.S.P. acknowledges that support for this work was provided by NASA through Hubble Fellowship grant \#HST-HF2-51379.001-A awarded by the Space Telescope Science Institute, which is operated by the Association of Universities  for  Research  in  Astronomy,  Inc.,  for  NASA,  under  contract  NAS5-26555. H.J. acknowledges the support of the Australian Research Council through Discovery Project DP150100862. 
The work of F.L. is supported by an ESO postdoctoral fellowship.
The Millennium-II Simulation databases used in this paper and the web application providing online access to them were constructed as part of the activities of the German Astrophysical Virtual Observatory (GAVO). The Illustris Simulation databases are provided by the Illustris Collaboration. The observational data we used are given in Table S1. The software for our Monte Carlo analysis is provided in Data S1, and the catalogues of Cen A analogues in the simulations are in Data S2 and S3.
The authors thank the anonymous referees for the helpful comments that improved the paper.\\

\textbf{Supplementary Materials}\\
www.sciencemag.org\\
Materials and Methods\\
Supplementary online text\\
Figs. S1, S2\\
Tables S1, S2\\
References (53-70)\\
Movie S1\\
Data S1, S2, S3\\
\newpage
\setcounter{figure}{0}
\captionsetup[figure]{name={Figure S},labelformat=nospace}
\captionsetup[table]{name={Table S},labelformat=nospace}
\renewcommand\theequation{S\arabic{equation}}
\title{\centering{\Huge Supplementary Materials for}\\
\vspace{0.5cm}
 {\Large A whirling plane of satellite galaxies around Centaurus A challenges cold dark matter cosmology\\}}
\author{{\centering Oliver M\"uller, Marcel S. Pawlowski, Helmut Jerjen, Federico Lelli\\}
{\centering correspondence to: oliver89.mueller@unibas.ch}}
\date{}


\baselineskip24pt


\maketitle

%
%________________________________________________________________
\section*{This PDF file includes:}
\begin{description}
\item {Materials \& Methods}
\item {Supplementary Text}
\item {Figs. S1 to S2}
\item {Tables S1 to S2}
\item {Caption for Movie S1}
\end{description}
\section*{{Other Supplementary Materials for this manuscript includes the following:}}
\begin{description}
\item {Movie S1}
\item {Data S1, S2, S3}
\end{description}
\newpage
\section*{Materials \& Methods}
The coordinates, distances and heliocentric velocities of the galaxies used in this work are compiled in Table\,S\ref{table:sample}. The coordinates and distances are adopted from \cite{2015ApJ...802L..25T,2016A&A...595A.119M}. The heliocentric velocities are taken from the online version\footnote{last checked: 11 December 2017} of the Local Volume catalog \cite{2004AJ....127.2031K,2013AJ....145..101K}, a compilation of objects within 11\,Mpc from the Local Group \cite{2015ApJ...802L..25T}. The main source of distance estimates are TRGB measurements using the Hubble Space Telescope. This method uses the sharp upturn in the stellar luminosity function produced by the red giant stars leaving the red giant branch during the explosive onset of the helium burning phase of evolution. The velocities were measured using either the 21-cm hydrogen emission line (H~\textsc{i} line) as part of the blind H~\textsc{i} Parkes All Sky Survey (HIPASS) \cite{2001MNRAS.322..486B}, or via dedicated optical spectroscopy of individual targets using various absorption lines (e.g. Balmer lines and Ca~\textsc{ii}).
 
\subsection*{General test of kinematic coherence for a planar satellite distribution}
\begin{figure}[H]
   \centering
   \includegraphics[width=12cm]{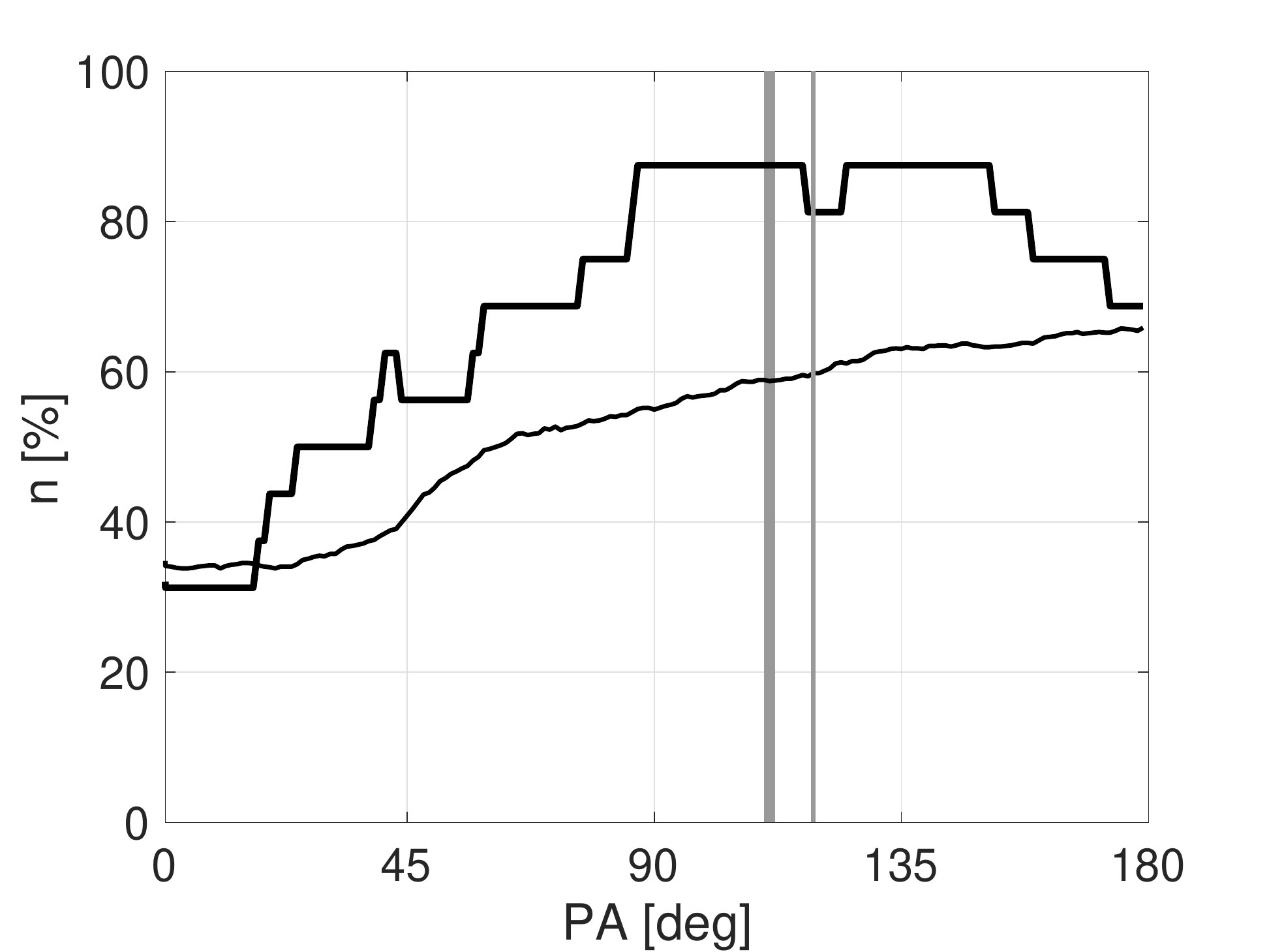}
   \caption{\textbf{Alignment of the kinematic coherence.} The fraction of approaching/receding satellite galaxies (thick black line) and planetary nebulae (thin black line) with respect to a separation line with variable position angle (PA) centered on Cen\,A.
The thick vertical gray line indicates the position angle of Cen\,A dust lane ($PA=110^\circ$). The thin gray lines is the position angle ($PA=119^\circ$) of the satellite plane projected minor axis.
}
\label{separation}
\end{figure}

If the heliocentric velocities of satellite galaxies are related to their planar distribution, the split between approaching and receding satellites should be maximal when the separation line is equal to the geometrical minor axis of the plane. To test this hypothesis we separate the satellite population with dividing lines centered on Cen\,A and $PA$ in the range $0^\circ-180^\circ$ (North to East).
Fig.\,S\ref{separation} shows that the number of satellite galaxies with coherent velocities is highest (14 coherent moving satellites out of 16, or 88$\%$) when the separation line corresponds to the geometrical minor axis of the Cen\,A system and smallest when it is close to the major axis ($0^\circ<PA<20^\circ $ and $175^\circ<PA<180^\circ$). The maximum coherence is actually achieved within two broad $PA$ ranges ($85^\circ<PA<110^\circ $ and $125^\circ<PA<155^\circ$). There is no a-priori reason why these intervals should include the geometric minor axis of the plane. This demonstrates that both satellite positions and velocities are consistent with a co-rotating plane. We performed the same test with the available kinematic data for planetary nebulae and found agreement within 5\,percent.

\subsection*{Tests of correlation between LoS velocities and separations to Cen\,A}
To explore the relation between the line-of-sight velocities of satellite galaxies and their separations to Cen\,A, we applied three different correlation tests to the data: the Pearson's R test, the Spearman's Rho test, and the Kendall's Tau test.  The resulting correlation coefficient is a number between $-1$ and 1, where $\pm$1 indicates a perfect positive or negative correlation, while 0 means no correlation. 
We used the algorithms implemented in the statistics toolbox of Matlab.
The tests were applied on the velocities and projected/3D separations of all galaxies. Every test finds a correlation within a 95\% confidence level. The correlation values (and corresponding $p$-values) for the projected separations are -0.599 (0.011), -0.529 (0.031), and -0.382 (0.034), respectively. For the 3D separation we find correlation within a 99\% confidence level, the values are  are -0.682 (0.003), -0.618 (0.010), and -0.485 (0.006), respectively.
Overall, there is strong evidence for correlated motion.

\subsection*{Monte Carlo simulations  of the  kinematic coherence}
\begin{figure}[H]
   %\centering
   \includegraphics[width=7.5cm]{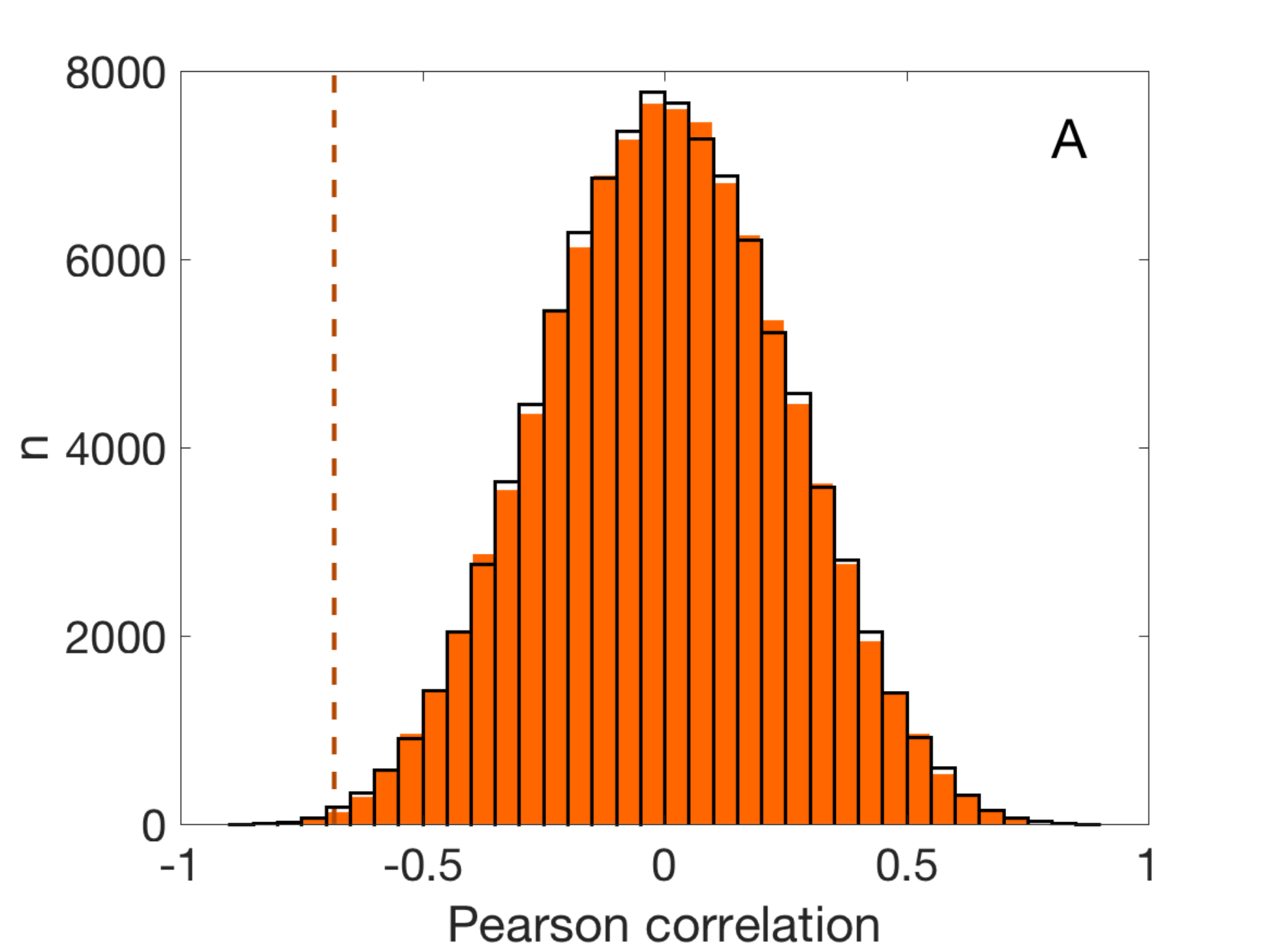}
   \includegraphics[width=7.5cm]{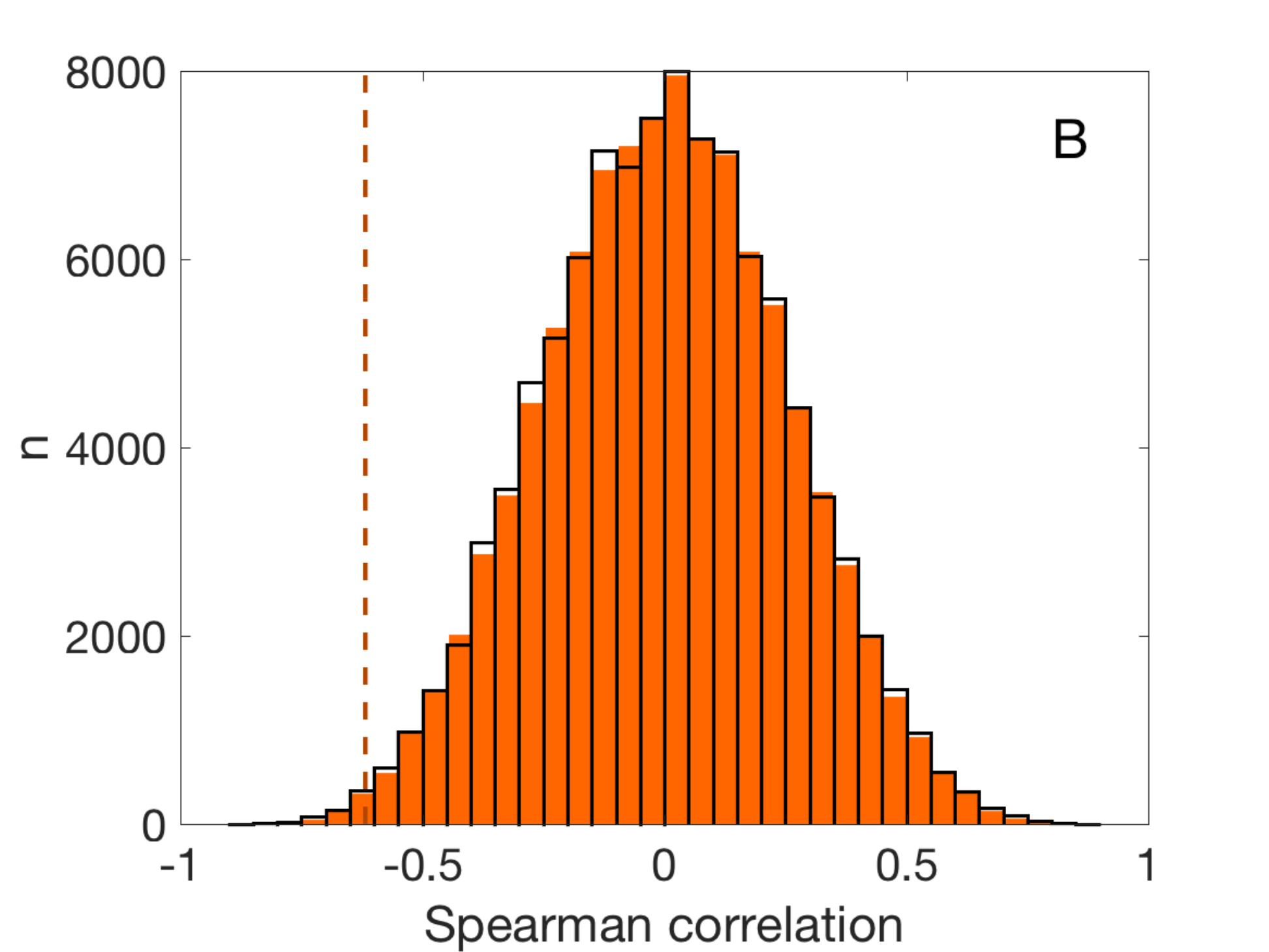}
   \includegraphics[width=7.5cm]{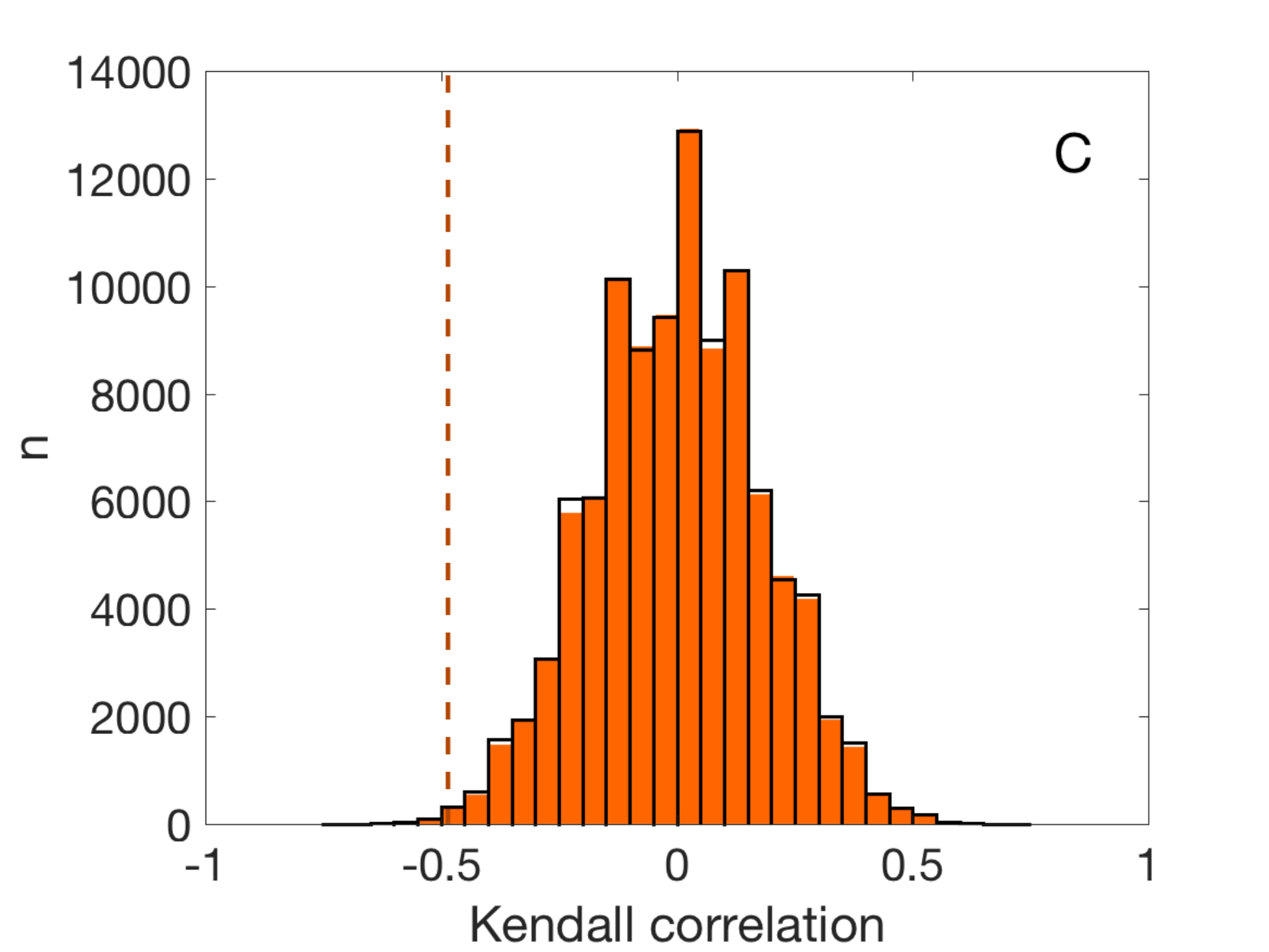}      \caption{\textbf{Monte Carlo simulations.} Results from Monte Carlo simulations for the three different correlation tests without (orange bins) and with (black line) considering the distance uncertainties. A correlation value of $\pm1$ corresponds to fully correlated data in positive or negative direction, respectively. The vertical dashed lines correspond to the measured correlation in the observed data. The bin width is 0.05.}
         \label{corr}
   \end{figure}

To further assess the kinematic coherence, we performed Monte Carlo simulations where we shuffled the measured velocities and randomized the sign of the 3D separation. Hence, every galaxy is assigned a new but measured velocity value. Its relative position to Cen\,A (north or south) is decided by a fair coin flip. As we measure the separation of Cen\,A to its satellites, this coin flip corresponds to randomizing the angles and keeping the radius fixed (in 50\% of the cases the satellite will lie to the north of Cen\,A, in 50\% to the south). On this new dataset, the three correlation tests were applied. This was repeated 100,000 times. The measured $p$-values follow uniform distributions, meaning that there is no favorite setup for correlated satellites. Figure S\ref{corr}\ shows a histogram of all correlation values in the Monte Carlo simulation. The correlations follow normal distributions with the mean around 0 (= no correlation). Our observed correlation values lie in the 3$\sigma$ tail. \\
We repeated this Monte Carlo method for the projected separation and confirm the previous results. As we used fixed distances we further investigated how the distance uncertainties affect our results by repeating the test, but this time randomizing the distance within a normal distribution (mean $\mu=D$ and standard deviation $\sigma=5$\% uncertainty) for every run. This again confirms our prior results.

\subsection*{Bayesian analysis }
{$P$-values smaller than 0.01 indicate a small chance of finding such a correlation in random, normal distributed data. We therefore consider how much more likely the hypotheses of correlated data is in respect to the null hypothesis of uncorrelated data. The Bayes Factor $BF$ quantifies the evidence of a model $M_1$ in favor of an alternative model $M_0$.
Here $M_1$ will correspond to correlated data, $M_0$ to uncorrelated data. The Bayes Factor is \cite{Wetzels2012}:
\begin{equation}
BF=(n/2)^{1/2}\cdot\Gamma(1/2)^{-1}\cdot 
\int_0^{\infty} (1+g)^{(n-2)/2}[1+(1-r^2)g]^{-(n-1)/2}g^{-3/2}\mathtt{e}^{-n/(2g)}dg
\end{equation}

where $n$ is the number of data points,  $\Gamma$ is the Gamma function, $g$ is the $g$-prior, and $r$ is Pearson's correlation value. A value of $BF$ larger than 1 favors the model $M_1$, otherwise it favors the model $M_0$. This numerical integration gives $BF=4.53$ and $BF=16.56$, respectively, meaning that with the given data the model $M_1$ is more likely than $M_0$, hence coherent moving satellites are indeed the statistically-favoured model.
}

\subsection*{Test of circular orbits within the satellite plane}
We tested whether the satellite galaxies are on circular orbits within the plane. In such circumstance, the circular velocity $V_{\rm c}$ of the satellite is related to its line-of-sight velocity $V_{\rm LoS}$ via the following equation:
\begin{equation}
V_{\rm LoS} = V_{\rm Cen\,A} + V_{\rm c}\sin(i)\cos(\theta)
\end{equation}
where $V_{\rm Cen\,A}$ is the systemic velocity of Cen\,A, $i$ is the inclination of the plane with respect to the sky, and $\theta$ is the azimuthal angle of the satellite within the plane. The azimuthal angle can be easily estimated by choosing a face-on orientation for the satellite plane.

The circular velocities of the satellites are expected to either decrease with distance from Cen\,A (like planets in the Solar System) or to reach a constant value (like gas and stars within galaxies). Instead, we find that the values of $V_{\rm c}$ vary randomly from galaxy to galaxy, suggesting that the orbits cannot be circular. Varying the distance of satellites within the uncertainties does not improve the result, hence we conclude that the orbits must be elliptical.

\subsection*{Comparison to $\Lambda$CDM simulations}

To determine how common the Cen\,A's satellite system is in $\Lambda$CDM simulations, we compare to two publicly available simulations: the dark-matter-only Millennium II simulation \cite{2009MNRAS.398.1150B} and the hydrodynamical Illustris simulation \cite{2014Natur.509..177V, 2015A&C....13...12N} which includes prescriptions for gas physics, star formation, and feedback processes. Specifically, for Millennium\,II we adopt the redshift zero galaxy catalogue \cite{2013MNRAS.428.1351G} which re-scales the simulation to Wilkinson Microwave Anisotropy Probe 7 (WMAP)  cosmological parameters, while for Illustris we use the redshift zero catalog of the highest-resolution Illustris-1 run \cite{2015A&C....13...12N}. 
We select as possible host galaxies all dark matter halos with a virial mass in the range of $4.0\ \mathrm{to}\ 12.0\,\times\,10^{12}\,M_{\odot}$. This mass range is selected to be consistent with several different halo mass estimates for Cen\,A \cite{2006AJ....132.2424W, 2007AJ....134..494W, 2007AJ....133..504K, 2010AJ....139.1871W}. 
Cen\,A is an isolated galaxy: its closest massive neighbor is M\,83 which lies 1.1\,Mpc behind Cen\,A and $\sim$13$^\circ$ degrees away. 
Thus, to make sure our host halos are similarly isolated as Cen\,A, we reject all possible hosts which have another halo of mass $\geq\ 1.0\,\times\,10^{12}\,M_{\odot}$\ within a distance of 1.4\,Mpc. This leaves us with 222 (Millennium\,II) and 146 (Illustris) isolated host galaxies.

For each host, 10 randomly oriented sight-lines are chosen, and the host and its surrounding galaxies are placed at Cen\,A's distance of 3.68\,Mpc from the observer's point of view. We then mock-observed the galaxy systems from this orientation, by projecting the angular positions relative to the host as well as the line-of sight velocities. All galaxies within $12^\circ$\ and separated by less then 0.8\,Mpc from the host are recorded as satellite galaxies. This is independent of whether they are actually within the virial radius of the simulated host halos. Nevertheless, in the following we will refer to them  as satellites for simplicity. We also reject all satellites within $1^\circ$\ of the LoS to the host, since these would be unobservable in front of Cen\,A (the closest Cen\,A  satellite with a measured line-of-sight velocity is ESO324-024 at an angular distance of $\sim$1.6$^\circ$). Only satellites with an $r$-band magnitude of -9 or brighter are considered in Illustris. This avoids selecting dark sub-halos that did not form any stars. 
We rank these satellites by their peak virial mass (Millennium\,II) or $r$-band magnitude (Illustris), and select at random 16 satellites ($N_\mathrm{kine}$) out of the top 30 satellites ($N_\mathrm{sat}$). This is because  kinematics (line-of-sight velocities) are only known for 16 out of the $\sim$30\ Cen\,A satellites with measured distances \cite{2015ApJ...802L..25T}. To test whether this selection affects the results we also repeat the analysis while selecting only the top $N_\mathrm{sat} = 16$\ satellites.
For Millennium\,II, in all cases a sufficient number of satellites was found within the mock survey volume. This leaves 2220 realizations of satellite systems ($N_\mathrm{realizations}$). In Illustris, a few realizations do not contain a sufficient number of satellites and are excluded, such that $N_\mathrm{realizations} = 1441$ (if the top 30 satellites are chosen) and $N_\mathrm{realizations} = 1459$ (if the top 16 satellites are chosen) are included in the analysis out of 1460 generated systems.

To avoid the look-elsewhere-effect, we apply some simplified criteria which the simulated satellite system has to fulfill to be counted as comparably correlated to the observed Cen\,A  system. This effectively results in an upper limit on the frequency of satellite systems as correlated as that of Centaurus A, i.e., we under-estimate any tension with the $\Lambda$CDM simulation.
We measure the overall flattening of the 16 satellite system on the sky and its kinematic coherence. By considering only the overall two-dimensional (projected) flattening, we avoid uncertainties based on distance measurements and the possibility of there being two parallel planes (see \cite{2015ApJ...802L..25T,2016A&A...595A.119M} for discussion). We measure the flattening $b/a$ by finding the short and long axes of the distribution using the tensor of inertia method \cite{2015MNRAS.453.1047P} and calculating the root-mean-square extend of the satellites along these axes.
The kinematic coherence $N_{\mathrm{corr}}$\ is measured as for the observed system, along the direction defined by the long axis of the satellite distribution. Applying the algorithm to the 16 observed Cen\,A satellites with measured kinematics, we obtain $b/a = 0.52$\ and $N_{\mathrm{corr}} = 14$. 
The results are compiled in Table\,S\ref{table:lcdm}, and illustrated in Fig. 3. In the following we study the frequency of finding Cen\,A like analogues in the cosmological simulations as flattened as the observed system ($f_\mathrm{flat}$); as kinematically correlated as the observed system ($f_\mathrm{corr}$); and fulfilling both criteria simultaneously ($f_\mathrm{both}$).

For Millennium\,II, we find that 433 out of 2220 realizations contain satellite systems that are at least as flattened ($b/a \leq 0.52$) as Cen\,A on the sky ($f_\mathrm{flat} = 19.5$\ per cent) and 26 that are sufficiently kinematically correlated ($N_{\mathrm{corr}} \geq 14$; $f_\mathrm{corr} = 1.17$\ per cent). This is in line with earlier findings \cite{2012MNRAS.424...80P} indicating that satellite systems in $\Lambda$CDM simulations are somewhat, but not strongly, more correlated than perfectly isotropic systems (for which the frequency of equally strong kinematic correlation would be 0.42 per cent).
Only 2 out of our 2220 realizations are simultaneously sufficiently flattened and sufficiently kinematically correlated to match the observed Cen\,A satellite system ($f_\mathrm{both} = 0.09$\ per cent). This makes the Cen\,A system a $\geq 3.3\sigma$\ outlier, indicating that it is a rare exception in $\Lambda$CDM. This low frequency is comparable to those reported for the satellite planes around the Milky Way and Andromeda \cite{2014ApJ...784L...6I,2014MNRAS.442.2362P}.
Even if we pre-select only those simulated satellite systems which are at least as extremely flattened as the observed system, only 0.46 per cent of these (2 of 433) display a kinematic coherence at least as extreme as that observed for Cen\,A. This number is consistent with that expected from random velocities.

The Millennium II galaxy catalogue contains so-called orphan galaxies: objects that are tracked even after their host dark matter halo has been disrupted. Their positions may be unreliable. Excluding these objects from our analysis reduces the frequency of sufficiently flattened (to 289 or 13.0 per cent) and sufficiently kinematically coherent (to 17 or 0.77 per cent) systems. The frequency of realizations fulfilling both criteria simultaneously stays the same (2 or 0.09 per cent).

For Illustris, 207 out of 1441 realizations are as flattened as Cen\,A on the sky ($f_\mathrm{flat} = 14.4$\ per cent), which is lower than for Millennium II ($f_\mathrm{flat} = 19.5$\ per cent). On the other hand, Illustris results in slightly higher frequencies of kinematically correlated satellite systems (23 out of 1441 realizations, or $f_\mathrm{corr} = 1.6$\ per cent) and of systems fulfilling both the flattening and the correlation criteria (7 out of 1441 realizations, or $f_\mathrm{both} = 0.49$\ per cent). The sample sizes of satellite systems in both simulations are too small to decide whether these are genuine effects of the modelling of baryonic physics in the Illustris simulation, or simply stochastic fluctuations.

Two of the observed satellite galaxies have line-of-sight velocities that overlap with that of Cen\,A within the uncertainties: NGC\,4945 and NGC\,5011C. In the unlikely case that both satellite velocities are revised to a lower value while Cen\,A's velocity is also revised to a higher value, this could in principle reduce the number of satellites with coherent velocities from $N_\mathrm{corr} = 14$\ to $N_\mathrm{corr} = 12$. In that case, finding a similar coherence in the simulations is more likely ($f_\mathrm{corr} = 11.2$\ and $f_\mathrm{corr} = 11.5$\ per cent for Millennium\,II and Illustris, respectively). Consequently, the frequency of systems fulfilling both the flattening and the velocity coherence criteria is increased to $f_\mathrm{both} = 2.16$\ per cent for Millennium\,II and to $f_\mathrm{both} = 1.67$\ per cent for Illustris.

In the previous analysis we randomly selected 16 out of 30 top-ranked satellites
to mimic the fact that velocities are only known for a subset of the confirmed Cen\,A satellites. To test this selection, we have repeated the previous analysis by using only the 16 top-ranked satellites. The results are also shown in Table\,S\ref{table:lcdm}. The resulting frequencies 
do not differ substantially or systematically from those previously found. This indicates that our selection does not bias our results. 

\subsection*{Supplementary Text}

Hydrodynamic simulations model baryonic processes such as gas cooling, star formation, stellar and nucleosynthetic evolution, supernova and black hole feedback. These simulations have been used to address discrepancies between $\Lambda$CDM and observations on small-scales, which were first identified in dark-matter-only simulations \cite{2017ARA&A..55..343B}. Modelling of baryons most directly affect the inner regions of dark matter halos ($\lesssim$10 kpc), since this is where most stars form and eventually explode as supernovae, injecting energy in the surrounding medium. The overall distribution and motion of satellite galaxies concern much larger spatial scales ($>100$ kpc), thus they are less directly affected by baryonic effects. However, these may not be entirely negligible. Baryonic effects can change the halo potential relative to a dark-matter-only case if they, for example, lead to the formation of a dark matter core. The formation of a central disk galaxy and resulting potential can furthermore enhance the tidal stripping of satellites. This can bias the distribution of satellite galaxies since the innermost satellites are preferentially destroyed, resulting in radially more extended satellite systems \cite{2017MNRAS.471.1709G}. Satellites on more radial orbits are expected to be more affected by tidal stripping, such that the orbital properties of the whole satellite system can also be affected.

In our analysis we find agreement with this tendency of more radially extended satellite systems for hydrodynamical simulations. Even though the same selection cuts are applied (satellites between $1^\circ$\ and $12^\circ$\ from their host, and within 800\,kpc), the  Cen\,A analog systems have an average root-mean-square radial extent $\langle R_\mathrm{rms}\rangle = 5.1^\circ$\ in the Millennium\,II simulation but are more extended in the Illustris simulation with an average $\langle R_\mathrm{rms}\rangle = 5.9^\circ$. The latter is close to the observed Cen\,A system, for which we measure $R_\mathrm{rms} = (6.1 \pm 0.8)^\circ$ with the uncertainty estimated via bootstrap resampling.

The overall flattening and kinematics of satellite systems, however, do not show substantial differences between the dark-matter-only Millennium\,II and the hydrodynamic Illustris simulation. Hence, we conclude that there is no evidence that baryonic effects are sufficient to result in a substantially increased fraction of extreme satellite planes comparable to the observed Cen\,A system.

\newpage

\newpage
\newpage

\begin{table}[H]
\caption{\textbf{Data used in this work.} Members of the Cen\,A subgroup \cite{2004AJ....127.2031K,2013AJ....145..101K,2015ApJ...802L..25T,2016A&A...595A.119M} with known distances and radial velocities.  For KKs\,59 we adopted the same distance as Cen\,A because there is no accurate TRGB distance available. (a): galaxy name, (b): right ascension in epoch J2000, (c): declination in epoch J2000, (d) galaxy distance, (e): reference for the distance measurement, (f): galaxy heliocentric velocity, and (g): reference for the velocity measurement.}             % title of Table
\label{table:sample}      % is used to refer this table in the text
\centering                          % used for centering table
\begin{tabular}{l c c l l l l l}        % centered columns (4 columns)
\hline\hline                 % inserts double horizontal lines
Galaxy Name & $\alpha_{2000}$ & $\delta_{2000}$ & $D$ & Reference & $v_h$ & Reference \\    % table heading 
 & (degrees) & (degrees) &  (Mpc) &  &(km\,s$^{-1}$) & \\    % table heading 
(a) & (b) & (c)& (d) & (e) & (f) & (g) \\    % table heading 
\hline      \\[-2mm]                  % inserts single horizontal line
ESO\,269-037        &         195.8875&       $-$46.5842&   	3.15$\pm$0.09& \cite{2004AJ....127.2031K,2013AJ....145..101K} &744$\pm2$& \cite{2004AJ....127.2031K,2013AJ....145..101K}\\
NGC\,4945           &         196.3583&       $-$49.4711&       3.72$\pm$0.03& \cite{2015ApJ...802L..25T} &563$\pm3$& \cite{2004AJ....128...16K}\\ %HIPASS 3.47
ESO\,269-058        &         197.6333&       $-$46.9908&       3.75$\pm$0.02& \cite{2004AJ....127.2031K,2013AJ....145..101K} &400$\pm18$& \cite{1999ApJ...524..612B}\\ %HIPASS
ESO\,269-066        &         198.2875&       $-$44.8900&       3.75$\pm$0.03& \cite{2004AJ....127.2031K,2013AJ....145..101K} &784$\pm31$& \cite{2000AJ....119..166J}\\
NGC\,5011C          &         198.2958&       $-$43.2656&       3.73$\pm$0.03& \cite{2004AJ....127.2031K,2013AJ....145..101K} &647$\pm96$& \cite{2007AJ....133.1756S}\\
KK\,196             &         200.4458&       $-$45.0633&       3.96$\pm$0.11&\cite{2004AJ....127.2031K,2013AJ....145..101K} &741$\pm15$& \cite{2000AJ....119..593J}\\
NGC\,5102           &         200.4875&       $-$36.6297&       3.74$\pm$0.39& \cite{2015ApJ...802L..25T}&464$\pm18$& \cite{2005MNRAS.361...34D}\\ %HIPASS 3.66
\it Cen\,A           &    \it 201.3667&\it    $-$43.0167&\it    3.68$\pm$0.05 & \cite{2004AJ....127.2031K,2013AJ....145..101K} & \it 556$\pm$10& \cite{2004AJ....128...16K}\\
ESO\,324-024        &         201.9042&       $-$41.4806&       3.78$\pm$0.09& \cite{2004AJ....127.2031K,2013AJ....145..101K}&514$\pm18$& \cite{2005MNRAS.361...34D} \\ %HIPASS
NGC\,5206           &         203.4292&       $-$48.1511&       3.21$\pm$0.01& \cite{2004AJ....127.2031K,2013AJ....145..101K} &583$\pm6$& \cite{1993AJ....105.1411P}\\
NGC\,5237           &         204.4083&       $-$42.8475&       3.33$\pm$0.02& \cite{2004AJ....127.2031K,2013AJ....145..101K}  &361$\pm4$& \cite{2004AJ....128...16K} \\ %HIPASS
NGC\,5253           &         204.9792&       $-$31.6400&       3.55$\pm$0.03& \cite{2015ApJ...802L..25T} &407$\pm3$& \cite{2004AJ....128...16K}\\ %HIPASS 3.44
KK\,211             &         205.5208&       $-$45.2050&       3.68$\pm$0.14& \cite{2004AJ....127.2031K,2013AJ....145..101K} &600$\pm31$& \cite{2008ApJ...674..909P} \\
ESO\,325-011        &         206.2500&       $-$41.8589&       3.40$\pm$0.05& \cite{2004AJ....127.2031K,2013AJ....145..101K} &544$\pm1$& \cite{2012MNRAS.420.2924K}\\ %HIPASS
KK\,221             &         207.1917&       $-$46.9974&       3.82$\pm$0.07& \cite{2004AJ....127.2031K,2013AJ....145..101K} &507$\pm13$& \cite{2008ApJ...674..909P}\\
ESO\,383-087        &         207.3250&       $-$36.0614&       3.19$\pm$0.03& \cite{2004AJ....127.2031K,2013AJ....145..101K} &326$\pm2$& \cite{2004AJ....128...16K} \\ %HIPASS
KKs\,59                 &         206.9920&       $-$53.3476&       3.68*&&686$\pm1$& \cite{2012MNRAS.420.2924K}\\
\hline
\end{tabular}
\end{table}

\newpage
\subsection*{Table S2}
\begin{table}[H]
\caption{\textbf{Comparison to $\Lambda$CDM simulations.} Frequencies of realizations of satellite systems in the Millennium\,II and Illustris simulations being as flattened ($f_\mathrm{flat}$) and as kinematically correlated ($f_\mathrm{corr}$) as the observed Cen\,A system, or fulfilling both criteria simultaneously ($f_\mathrm{both}$).}             % title of Table
\label{table:lcdm}      % is used to refer this table in the text
\centering                          % used for centering table
\begin{tabular}{l c c l l l}        % centered columns (4 columns)
\hline\hline                 % inserts double horizontal lines
 & $N_\mathrm{realizations}$ & $N_\mathrm{kine}/N_\mathrm{corr}$ & $f_\mathrm{flat}$ & $f_\mathrm{corr}$ & $f_\mathrm{both}$\\    % table heading 
Simulation sample &  &  & (\%) & (\%) & (\%)\\    % table heading 
\hline      \\[-2mm]                  % inserts single horizontal line
% all results for satellites with r-band magnitude of -9 or brighter
%
16 out of top 30 satellites \\
Millennium\,II     &  2220 &  14/16 &   19.5 &  1.17 &  0.09 \\
Illustris          &  1441 &  14/16 &   14.4 &  1.60 &  0.49 \\
Millennium\,II     &  2220 &  12/16 &   19.5 &  11.2 &  2.16 \\
Illustris          &  1441 &  12/16 &   14.4 &  11.5 &  1.67 \\
\hline      \\[-2mm]                  % inserts single horizontal line
Top 16 satellites \\
Millennium\,II     &  2220 &  14/16 &   17.5 &  0.50 &  0.18 \\
Illustris          &  1459 &  14/16 &   15.9 &  1.30 &  0.27 \\
Millennium\,II     &  2220 &  12/16 &   17.5 &  10.7 &  2.34 \\
Illustris          &  1459 &  12/16 &   15.9 &  13.6 &  2.26 \\
\hline
\end{tabular}

\end{table}
\newpage

\subsection*{Caption for Movie S1}
The movie starts by showing the 3D spatial distribution of all confirmed satellites of Cen\,A (black dots) and the candidate members (open circles) with their predicted distances \cite{2016A&A...595A.119M} in equatorial cartesian coordinates. The dashed line is our line-of-sight towards Cen\,A. The plane of satellites (gray) is faded in. Then, the satellites without measured velocities are faded out and the kinematic information of the satellites is presented by a color and a line (red and blue for receding and approaching, respectively). 
 The length of the lines is proportional to the observed velocity.

\end{document}